
\input harvmac
\let\<=\langle 
\let\>=\rangle

\def\l{\lambda}
\def\D{\Delta}
\def\p{\phi}
\def\det{\hbox{det}}
\def\N{^{(N)}}
\def\bt{\bar t}
\def\dim{\hbox{dim}}

\def\Tr{\hbox{Tr}}
\def\cor{\< \p _{i_1}\cdots \p _{i_n}\> _g}
\pretolerance=750 
 
\Title{\vbox{\baselineskip12pt\hbox{HUTP-91/A041}}} 
{\vbox{\centerline{Fusion Residues}}} 
\centerline{Kenneth Intriligator} 
\bigskip\centerline{Lyman Laboratory of Physics} 
\centerline{Harvard University}\centerline{Cambridge, MA 02138}

\vskip .3in 
  
We discuss when and how the Verlinde dimensions of a rational conformal field
theory can be expressed as correlation functions in a topological LG theory. 
It is seen that a necessary condition is that the RCFT fusion rules must
exhibit an extra symmetry.   We consider two particular perturbations of the
Grassmannian superpotentials.  The topological LG residues in one
perturbation, introduced by Gepner, are shown to be a twisted version of the
$SU(N)_k$ Verlinde dimensions.  The residues in the other perturbation are the
twisted Verlinde dimensions of another RCFT; these topological LG correlation
functions are conjectured to be the correlation functions of the
corresponding Grassmannian topological sigma model with a coupling in the
action to instanton number. 
\vskip 1in

\Date{8/91}

 
\newsec{Introduction} 
      
We can associate with any rational conformal field theory a topological
theory whose correlation functions are the dimensions of the Friedan-Shenker
\ref\FS{D. Friedan and S. Shenker, Nucl. Phys. B281 (1987) 509}
vector bundles:
\eqn\fs{\cor ^o=\dim {\cal H}(g;i_1,\dots ,i_n)}
where ${\cal H}(g,i_1,\dots ,i_n)$ is the space of conformal blocks for the
genus $g$ surface with insertions of the primary fields $\p _{i_1},\cdots \p
_{i_n}$ (we will consider a related,``twisted'', version of these correlation
functions, hence the superscript).  These topological integers can be
expressed in terms of the modular
transformation matrix $S$ as:
\eqn\v{\cor ^o= \sum _p {1\over {S_{op}}^{2(g-1)}}{S_{i_1p}\over S_{op}}\cdots
{S_{i_np}\over S_{op}}}
where $p$ runs over all primary fields \nref\ver{E. Verlinde, Nucl. Phys. B300
(1988) 360}\nref\ms{G. Moore and N. Seiberg, {\it Lectures on RCFT} in
{\it Superstrings '89} (Trieste proc.) M. Green et. al. (Eds.), World
Scientific, and in {\it Physics, Geometry, and Topology} (Banff '89 proc.)
NATO ASI B238, H.C. Lee (Ed.), Plenum (1990)}\refs{\ver ,\ms}.  This
expression is of
the expected topological form:
$$\cor =\Tr h^{g-1}\p _{i_1}\cdots \p _{i_n}$$
where $h$ is the operator which makes a handle \ref\w{E. Witten, Nucl. Phys.
B340 (1990) 281}.    The Verlinde dimensions \fs\ in $G_k$ theories can be
understood as Chern-
Simons amplitudes on $\Sigma _g\times S ^1$ \ref\cs{E. Witten, Commun. Math.
Phys. 121 (1989) 351} or as amplitudes in $G/G$ theory \nref\sp{M. Spiegelglas,
{\it Fusion Rules as Amplitudes in }$G/G$ {\it Theories}, TECHNION-
PH-35-90}\nref\hf{E. Witten, {\it On Holomorphic Factorization of WZW and
Coset Models}, IASSNS-HEP-91/25}\refs{\sp ,\hf}.

The expression \v\ 
is similar to the residue expression for correlation functions in topological
LG theories \ref\tlg{C. Vafa, Mod. Phys. Lett. A6 (1991) 337}.  The
correlation function at genus $g$ of an arbitrary function $F(X_i)$ of the
chiral superfields $X_i$ $(i=1,\dots ,N)$ in a topological LG theory with
superpotential $W(X_i)$ is given by \tlg 
\eqn\res{\< F\> _g=\sum _{dW=0}h^{g-1}F \quad \hbox{where} \quad
h=(-1)^{N(N-1)/2}\det (\partial
_i\partial _j W).}  
We have here modified the operator $h$ from that
obtained in \tlg\ to include a convenient sign.  This sign naturally arises in
topological LG theory from the fermion zero mode integration upon ordering the
terms in the Grassmann integrals so that the holomorphic pieces are all
adjacent to each other.  In this note we will
study the connection between \v\ and the residue expression \res\ of
topological LG theory. 

The idea is to find a topological LG superpotential $W(X)$ with chiral primary
fields $\p _i(X)$ so that $\p _i$ at the $p$-th critical point (the $p$-th
solution of $dW=0$) is $\p
_i(p)=S_{ip}/S_{op}$.  From \ver , $\p
_i(p)=S_{ip}/S_{op}$ form diagonal representations of the fusion rules for
every $p$.  This basis is the basis of critical points of $W$ where fusion is
diagonal ring multiplication.  The fusion rules are thus represented by
multiplication in the ring with ideal generated by $dW=0$.  This was recently
accomplished by Gepner \ref\g{D.
Gepner, {\it Fusion Rings and Geometry} preprint NSF-ITP-90-184} for the case
of $SU(N)_k$ theories. 

The polynomial $W(X)$ whose ideal $dW=0$ gives
fusion rule relations for some rational conformal field theory must be a
perturbed superpotential i.e. it can not
be quasihomogeneous.  This is clear since a quasihomogeneous superpotential
has a conserved $U(1)$ charge which is positive for all elements in the chiral
primary ring, additive under ring multiplication, and no element in the ring
has charge greater than $\hat c=c/3$ \ref\lvw{W. Lerche, C. Vafa, N. Warner,
Nucl. Phys.
B324 (1989) 427}.  Quasihomogeneous superpotentials
thus always have nilpotent ring relations whereas fusion rules in conformal
field theory, by duality, can never have zero on the RHS.  Our fusion rule
polynomial
$W(X)$ is thus a perturbation of a quasihomogeneous superpotential $W^o(X)$ in
some of the chiral ring elements $\p ^o_i$.  Under this perturbation, $\p
^o_i\rightarrow \p _i$ and the structure constants $c_{ij}^{o\ k}$ for ring
multiplication of the chiral primaries become the fusion rules $N_{ij}^k$.

We can now say something about the type of fusion rules which can be described
as above as ring multiplication associated with some perturbed superpotential.
As discussed in \ref\dvv{R. Dijkgraaf, E.
Verlinde, and H. Verlinde, Nucl. Phys. B352 (1991)
59},
the topological metric
\eqn\fumetric{\eta _{ij}=\< \p _i \p _j\> _{g=0}=N_{ij}^k\< \p _k\> _{g=0}}
is unchanged from that of the unperturbed theory.  We thus have
\eqn\fumetricu{\eta _{ij}=c_{ij}^{o\ k}\< \p ^o_k\> _{g=0}=c_{ij}^{o\ \hat
c}=N_{ij}^{\hat c}}
since $\p ^o_{\hat c}$, the unique chiral primary with the maximal $U(1)$
charge of $\hat c$ (it generates one unit of spectral flow), is the only
chiral primary with nonzero one point function on the sphere \tlg\ (and we can
choose our normalizations to make its one point function one) and the same
holds in the perturbed theory.  In any unperturbed
LG theory it is natural and always possible to choose our basis of $\p ^o_i$
so that $c_{ij}^{o\ \hat c}=\delta _{ \tilde \imath ,j}$ where the chiral
primary
$\p ^o_{\tilde \imath}$ is given by applying $\p _{\hat c}$ (one unit of
spectral
flow) to the anti-chiral primary dual to $\p ^o_i$, i.e. $\p ^o_{\tilde
\imath}=\p
^o_{\hat c}{\bar \p} ^o_{\bar \imath}$.  We thus naturally have $N_{ij}^{\hat
c}=\delta _{\tilde \imath , j}$.  From this it follows that the fusion rule
for the
field $\p _{\hat c}$ with any primary field $\p _i$ satisfies
\eqn\sfbs{\p _{\hat c}\times \p _{i}=\p _{i ^{\prime}}}
with only one term on the RHS.  Such fields were
considered in \nref\k{K. Intriligator, Nucl. Phys. B332 (1990)
541}\nref\sy{A.N.
Schellekens and S. Yankielowicz, Nucl. Phys. B327 (1989) 673}\refs{\k ,\sy}. 
They indicate
an extra symmetry in the theory above that of the chiral algebra \k .  We thus
expect
that only fusion rules with such fields (additional symmetries) can be
represented as ring
multiplication associated with some polynomial $W$.  In $G_k$ theories, the
center of $G$ is generated by fields that satisfy \sfbs \ref\fg{J. Fuchs and
D. Gepner, Nucl. Phys. B294 (1987) 30}.  The $(E_8)_3$ fusion rules obtained
in \ref\vDriel{J. Fuchs and P. van Dreil, Nucl. Phys. B346 (1990) 632}, for
example, contain no such field ($E_8$ has no center) and, therefore, can not
be naturally represented as multiplication in
the ring of some $W$.

Given a potential $W$ and chiral primary fields $\p _i$ with ring
multiplication the fusion rules of some conformal field theory, genus one
correlation functions computed with the residue expression \res\ of
topological LG theory will coincide with the genus one Verlinde dimensions \v\
since they both follow from just the fusion rules.  The genus zero correlation
functions computed with \res , on the other hand, differ slightly from the
corresponding Verlinde dimensions.  The residue formula, for example, gives
$\<\p _i\>_{g=0}=\delta _{i,\hat c}$ whereas \v\ gives  $\< \p _i \>
_{g=0}^o=\delta _{i,o}$ ($\p _o$ denotes
the
identity).  It is thus clear that the handle-making operator $h$ of
topological LG theory is a ``twisted'' version of that in the Verlinde formula
\v :
\eqn\th{h=h^o\p _{\hat c}.}
The untwisted Verlinde dimensions \fs\ can thus be computed as topological LG
residues by soaking up $1-g$ units of the spectral flow primary field $\p
_{\hat c}$ on the genus $g$ surface:
\eqn\soakres{\cor ^o=\< \p _{i_1}\dots \p _{i_n} (\p _{\hat c})^{1-g}\>
_g,}
where the correlation function on the right is computed in the
corresponding topological LG theory using the residue expression \res\ and
where $(\p _{\hat c})^{-1}$ is well defined by \sfbs\ and equal to the dual to
$\p _{\hat c}$.  The twisting \th\ is familiar in twisted N=2 theories; the
twisting
of the spins requires the handle-making operator to carry a unit of spectral
flow.  It would be interesting to understand this twisting in $G_k$
theory, where the twisting is by the center of $G$, in terms of Chern Simons
or $G/G$ theory.

We will apply these ideas to two particular perturbations of the Grassmannian
superpotentials of \lvw .  The plan for this paper is as follows.  In sect. 2
we
discuss the cohomology ring of the Grassmannian manifolds and the
superpotential
of \lvw\ which gives this ring.  We also discuss the perturbation of these
Grassmannian superpotentials in an interesting subset of the chiral primary
fields, generalizing somewhat some results of \dvv .  In sect. 3 we consider
the perturbed Grassmannian superpotentials found in \g\ to give the $SU(N)_k$
fusion rules.  We examine the residue
expression and explicitly verify \th\ in these theories.  The Verlinde
dimensions for $SU(N)_k$ can thus be expressed as residues, using \res , as in
\soakres .  In Sect. 4 we write the Verlinde dimensions of the $N=0$
unitary minimal models as TLG coset residues.  In sect. 5 we consider
another interesting perturbation of the Grassmannian superpotentials, the
perturbation in only the most relevant operator (the Kahler form).  The
correlation functions \res\ in this theory give the twisted Verlinde
dimensions \soakres\ for a RCFT.  It is conjectured that these correlation
functions are the correlation functions of a topological sigma model on the
Grassmannian manifold with a coupling in the action to instanton number.   
Sect. 6 is the conclusion.
We comment there about an apparent connection
between the integrability of a perturbed $N=2$ LG theory and having the
perturbed structure constants being the fusion rules of some rational
conformal field theory.  

\newsec{Grassmannian Superpotentials and Their Perturbations}

The Grassmannian $U(N+M)/U(N)U(M)$ (with, say, $N \leq M$) is a $NM$ complex
dimensional compact complex manifold
(a generalization of projective space) with a
cohomology ring that is generated by the forms $X_1,\dots , X_N$ where $X_r$
is a $(r,r)$ form \ref\btgh{R. Bott and L. W. Tu {\it Differential Forms in
Algebraic Topology}, Springer-Verlag (1982); P. Griffith and J. Harris {\it
Principles of Algebraic Geometry}, J. Wiley and Sons (1978); W. Stoll, {\it
Invariant Forms on Grassmann
Manifolds}, Annals of Mathematics Studies 89, Princeton University Press,
1977}.  Writing
\eqn\xtyt{X\N (t)=\sum _{i=0}^NX_it^i \quad \hbox{and}\quad Y\N (t)=(X\N
(-t))^{-1}=\sum _n Y\N _nt^n} 
with $X_0=1$,
the ideal of our Grassmannian's cohomology ring is obtained by the
requirements \btgh\ that
\eqn\yo{Y\N _i=0 \quad \hbox{for} \quad i=M+1,\dots , M+N.}
The Grassmannian cohomology elements are described in terms of the generating
$X_i$ as follows \btgh .  For every formal, $N$ dimensional vector $\mu = \sum
_{l=1}^N n_le_l$
with components $n_l$ that satisfy 
\eqn\nt{n(\mu )\equiv \sum _{l=1}^Nn_l\leq M \quad
n_l=0,1,2,\dots ,}
there is a $(r(\mu ),r(\mu ))$ form $\p _{\mu}$ in the
Grassmannian cohomology, where $r(\mu )=\sum _{l=1}^Nln_l$, given in terms of
the generating $X_i$ by
\eqn\grforms{\p _{\mu}\N =\det _{1\leq i,j \leq n(\mu )} X_{a_i+i-j}}
where the $a_i$ satisfy $a_{i}\leq a_{i+1}$ with $n_l$
of the $a_i$ equal to $l$ for each $l=1,\dots ,N$, and $X_r$ is defined to be
zero when $r<0$ or $r>N$ and $X_0=1$.  The $(N+M)!/N!M! $ forms $\p _{\mu}$
with
$\mu$ satisfying \nt\ are the Grassmannian cohomology elements.  In
particular, the $\p _{e_i}=X_i$ with $\p _{e_1}=X_1$ the Kahler form, and $\p
_{Me_N}=X_N^M$ is the volume form.  It is easily seen that the $\p _{ne_1}$
satisfy
$\p _{ne_1}=X_1\p _{(n-1)e_1}-X_2\p _{(n-2)e_1}+\dots $ from which it follows
that $\p \N _{ne_1}=Y \N _n(X_i)$ as defined by \xtyt .
From
\nt\ it is clear that the Grassmannian
Poincare polynomial is the generating function for the number of partitions of
a number into no more than $M$ parts with each part no greater than $N$:
\eqn\gpp{\prod _{i=1}^N{(1-(t\bt )^{M+i})\over (1-(t\bt )^i)}.}

We can represent the cohomology element $\p _{\mu}$ with a Young tableaux
containing $n_l$ columns of $l$ boxes for $l=1\dots N$.  Condition \nt\ says
that no tableaux has more than $M$ columns.  There is a duality under
$N$$\leftrightarrow $$M$, $X_i$$\leftrightarrow $$Y_i$, and in terms of the
tableaux, rows$\leftrightarrow$columns.  In particular, 
define $\hat n_l$ to be the number of rows in the $\p _{\mu}$ tableaux with
$l$ boxes for $l=1,\dots M$.  The elements \grforms\ can be expressed as:
\eqn\duality{\p _{\mu}=\det _{1\leq i,j \leq \hat n(\mu )}Y_{{\hat a}_i+i-j},}
where $\hat n(\mu )\leq N$ and the $\hat a_i$ are defined in terms of the
$\hat
n_l$ as in \grforms .  Writing $Y_r=\p _{re_1}$ in terms of the $X_i$ as in
\xtyt , it can be verified that \duality\ is equivalent to \grforms . 

The cohomology ring is obtained by multiplication of the elements \grforms\
subject to the ideal \yo .  Before imposing the relations of the ideal, the
product of any two elements \grforms\ can be expanded as a sum of terms of the
same type with nonnegative integer coefficients.  The ideal sets to zero those
terms in the sum which don't satisfy the
inequality \nt .  This ring structure is known as the Schubert calculus
\btgh . 
   
The ideal \yo\ of the Grassmannian cohomology ring can be written as the ideal
generated by setting to zero the derivatives of a polynomial in the $X_i$
 as follows.  Following \g , we define
\eqn\wt{W\N (t)=-\log (X\N (-t))=\sum _i W\N _it^i.}
Since
\eqn\Wderiv{{\partial W\N (t)\over \partial X_i}=-Y\N (t)(-t)^i \quad
i=1,\dots ,N,} 
our Grassmannian ideal \yo\ is generated by the derivatives of
$W\N _{N+M+1}(X_i)$.  If we write 
\eqn\xt{X\N (t)=\prod _{i=1}^N(1+q_it),}
it can be seen that
$$W\N _{N+M+1}(X_i)=\sum _{i=1}^N {q_i^{N+M+1}\over N+M+1}$$
written in terms of the $X_i$, the form obtained in \lvw .  Using
$W\N _{N+M+1}$ as a LG superpotential gives the $\hat c$=$c/3$=$NM/N$+$M$+1
Kazama Suzuki N=2 coset theory $SU(N+1)_{M}/SU(N)U(1)$.  The chiral primary
fields \grforms\ have $q$=$\bar q$=$(N+M+1)^{-1}r(\mu )$.  The operator $h$
defined in \res\ is proportional to the
top form $\p _{\hat c}=\p
_{Me_N}=X_N^M$.  The relation is
\eqn\hessprop{h={(N+M)!\over N! M!}\p _{\hat c}.} 
The residue expression
\res\ thus gives $\< \p _{\hat c}\> _{g=0}=1$ with our normalizations.   

We now consider deforming the Grassmannian superpotential 
in the chiral primary fields \grforms .  As discussed in \tlg , only
deformations in those $\p _{\mu}$ with $q<1$ are
nice in that they don't change the number of critical points of $W$.  Under
the perturbation $W(X_i)\rightarrow W(X_i,a)=W(X_i)+O(a)$ terms and $\p \N
_{\mu}(X_i) \rightarrow
\p \N _{\mu}(X_i;a)=\p \N _{\mu}(X_i)+O(a)$ terms such that 
\eqn\dvva{\eqalign{{\partial W\N (X_i;a)\over \partial a_{\mu}}&=-\p
_{\mu}(X_i;a)\cr
\p \N _{\mu _1}(X_i;a)\p \N _{\mu _2}(X_i;a)&=\sum _{\mu _3}c_{\mu
_1 \mu _2}^{\mu _3}(a)\p \N _{\mu _3}(X_i;a),\cr}}
where $c_{\mu _1 \mu
_2}^{\mu _3}(a)$ are the perturbed structure constants (the perturbed three
point functions with an index raised using the (a independent) metric $\eta $)
\dvv . 

The form of $W(X;a)$ and $\p _{ne_1}(X,a)$ in the one variable case, $N=1$,
was
obtained in \dvv .  We will generalize their results to the general $N$
variable case when the Grassmannian superpotential is perturbed in only the
chiral primary fields $\p \N
_{ne_1}=Y\N _n$ with $n=1,\dots M$ (note that these all satisfy the $q<1$
condition).
>From \Wderiv\ it is obvious that condition \dvva\ for these perturbations:
\eqn\Wpertderiv{{\partial W\N _{N+M+1}(X_i;a)\over \partial
a_n}=-Y\N _n(X_i;a)} 
is satisfied by :
\eqn\pert{\eqalign{W\N
_{N+M+1}(X_i;a)&=W^{(N+M)}_{N+M+1}(X_i,X_{N+n})\cr
\p \N_{\mu}(X_i;a)&=\p
^{(N+M)}_{\mu}(X_i,X_{N+n})\cr
\hbox{with}\ X_{N+n}&=(-1)^{N+n}a_{M+1-n}  \quad n=1,\dots M.\cr}}  
In particular, the perturbed $Y\N
_n(X_i;a)$ are given as in \xtyt\ where, now, 
$$X(-t;a)=\sum _{i=0}^NX_i(-t)^i+\sum _{n=1}^Ma_nt^{N+M+1-n}.$$ 
For the one variable (N=1) case these
results are
equivalent to those obtained in \dvv .  

The perturbed cohomology ring \dvva\ can be embedded in the unperturbed
Grassmannian cohomology ring in $N$+$M$ variables:
\eqn\pertring{\p _{\mu _1}^{(N+M)}\p _{\mu _2}^{(N+M)}=\sum _{\mu _3}\p _{\mu
_3}^{(N+M)}.}
While $\mu _1$ and $\mu _2$ are of the form $\sum _{l=1}^Nn_le_l$, with
components that satisfy \nt , the terms on the right are of the form $\mu
_3=\sum _{l=1}^{N+M}n_le_l$ (since this is now the $N$+$M$ variable Schubert
calculus).  It is easily seen from \grforms\ that the new terms with
$n_{l>N}\neq 0$ go away when the perturbation is turned off.  Using \grforms\
and the ideal of the
ring, obtained by setting to zero the $X_i$ derivatives ($i=1,\dots N$) of the
perturbed potential \pert , all terms on the RHS can be expanded out in terms
$n_{l>N}=0$ basis $\p \N _{\mu}(X_i;a)$ in \pert\ times powers of the
perturbing parameters $a$.  The perturbed ring structure in \dvva\ is thus
easily obtained.   

The topological LG correlation functions in the perturbed theory can be
calculated using \res .  It is sometimes convenient to work in terms of the
variables $q_i$ defined in \xt .  The Jacobian
between these variables is det$\partial q_i/\partial X_j=\Delta ^{-1}$ where
$\Delta$ is the Vandermonde determinant $\Delta =\prod _{i<j}(q_i-q_j)$.  The
critical points of $W$ are the critical points of $W$ in terms of the $q_i$
other than those points
$q_i=q_j$ where $(\Delta )^{-1}$ blows up and modulo permutations of the $q_i$
(this change of variables trick was used in \g\ and \ref\cv{S. Cecotti and C.
Vafa,
{\it Topological Anti-Topological Fusion}, HUTP-91/A031, SISSA-
69/91/EP}).  The operator $h$ in \res\ can be written up to the equations of
motion as $h=(-1)^{N(N-1)/2}\Delta ^{-2}\det(\partial ^2W/\partial q_i\partial
q_j)$. 

In the next two sections we will
consider two interesting special cases: perturbing in only the most relevant
operator $Y\N _1$=$\p \N _{\mu=e_1}$=$X_1$, and perturbing in only the least
relevant
operator of the type considered $Y\N _M=\p \N _{\mu=Me_1}$.         

\newsec{$SU(N)_k$ Fusion Residues}

In this section we consider a special perturbation of the Grassmannian
potentials: perturbation in only the least relevant operator of
the $\p _{ne_1}=Y_n$ type.  This perturbation was found by Gepner to give the
$SU(N)_k$ fusion rules.  This perturbation is also special in that it is
probably integrable \nref\pk{P. Fendley, K. Intriligator, in
preparation}\refs{\cv ,\pk} and its soliton structure is rather interesting
\pk .  It was
also shown in \ref\setting{M. Spiegelglas, {\it Setting Fusion Rings in
Topological Landau-Ginzburg}, TECHNION-PH-8-91} in the one variable
case (the $SU(2)_k$ case) that the perturbed ring structure constants
$c_{ij}^k(a)$ are symmetric in $i,j,k$ (as the $N_{ij}^k$ of $SU(2)_k$ are so
symmetric)
at precisely this perturbation. 
    
The polynomial $W(X)$ found by Gepner \g\ to give the fusion rules for
$SU(N)_k$ is given by
$W(X)=W^{(N)}_{N+k}(X_r,X_N=1)$, the Grassmannian polynomial
for $U(N+k-1)/U(N)U(k-1)$ subject to the constraint $X_N=1$.  This is the
perturbation \pert\ of the Grassmannian polynomial $W^{(N-1)}_{N+k}(X_1,\dots
, X_{N-
1})$ for $U(N+k-1)/U(N-1)U(k)$ in the chiral primary field $\p ^{(N-
1)}_{\mu=ke_1}$ with the perturbing parameter set to one.  
The Grassmannian chiral primaries $\p ^{(N-1)}_{\mu}$ with $\mu=\sum
_{r=1}^{N-1}n_re_r$
are given at the perturbed point by \pert , $\p _{\mu}\N (X_r,X_N=1)$.  This
is precisely the polynomial found
by Gepner \g\ to represent the $SU(N)_k$ primary field with highest weight
$\mu =\sum _{r=1} ^{N-1} n_r\Lambda _r$.  The Grassmannian
structure constants at this perturbed point are the $SU(N)_k$ fusion rules of
\ref\gw{D. Gepner and E.
Witten, Nucl Phys. B278 (1986) 493}.   
The $X_r$, in particular, represent the fundamental weights $\Lambda _r$, the
fully antisymmetric representations.  The $j$-th fully symmetric
representation, $\mu = j\Lambda _1$, is represented by
the polynomial $Y_j^{(N)}(X_r,X_N=1)$ \xtyt .  The field $\p _{\hat c}=\p
_{k\Lambda _{N-1}}$ satisfies the fusion rule \sfbs .  It generates the ${\bf
Z}_N$ center of $SU(N)$.

We will now consider the residue formula \res\ 
for
topological amplitudes using the above superpotential.  As discussed in the
introduction we expect to get \th\ 
\eqn\sunh{h=h^o\phi _{k\Lambda _{N-1}},}
from which it follows that the
residue
expression of topological LG theory gives a twisted version of the Verlinde
dimensions (where the twisting is by the center of the group).  We will
explicitly verify \sunh\ by computing \res\  
\eqn\sunhess{h=(-1)^{(N-1)(N-2)/2}\det ({\partial ^2W^{(N-
1)}_{N+k}(X_r;a)\over \partial X_r\partial X_s})}
where, again,
$W^{(N-1)}_{N+k}(X_r,a)=W\N _{N+k}(X_r,X_N=1)$.
  
Following Gepner, we re-introduce the field $X_N$ along with a Lagrange
multiplier to set it to one:
\eqn\Wl{W(X_r,X_N,\l )=W_{N+k}(X_r,X_N)-\l (X_N-1)=\sum _{l=1}^N
{q_l^{N+k}\over N+k}
-\l (\prod _{l=1}^Nq_l -1).}
The equation of motion of the polynomial \Wl\ for $\l$ sets
$X_N=1$ and, using \wt , the equation of motion for $X_N$ gives, 
$\l = (-1)^{N+1}Y_k(X_r)$.  It is easily seen that \sunhess\ is given by
$h=-(-1)^{(N-1)(N-2)/2}\det H$, where $\det H$ is the Hessian of $W(X_r,X_N,\l
)$.
In terms of the $q_i$, we have $\det H=(\Delta )^{-2} \det K$
where $K_{i,j}=\partial ^2 W/\partial q_i\partial q_j$, $K_{i,\l}=-q_i$, 
$K_{\l , \l}=0$ for $i,j=1,\dots ,N$ and, recall, $\D $ is the
Vandermonde determinant Jacobian. 
Using the
equations of motion $q_i^{N+k}=\l =(-1)^{(N+1)}Y_k$ and $\prod _{i=1}^N
q_i=1$, we obtain $h=-(-1)^{N(N-1)/2}\D
^{-2}(Y_k)^{N-1} \det A$,
where $A_{i,j}=(N+k)\delta _{i,j}$, $A_{i,\l }=1$, 
$A_{\l , \l}=0$ and, thus, $\det A=-N(N+k)^{N-1}= -|W/(N+k)R|$, the index of
$(N+k)$ times the $SU(N)$ root lattice in the $SU(N)$ weight lattice.  Since
$(Y_k)^{N-1}=\p _{k\Lambda _{N-1}}$ in the ring, we obtain
\eqn\hw{h=(-1)^{N(N-1)/2}\big| {W\over (k+N)R }\big| \D ^{-2}\p
_{k\Lambda _{N-1}}.}
In \g\ it was shown how to make the correspondence between weights $\mu$ of
$SU(N)_k$ and the critical points $X_r^{(\mu )}$ of $W$.  It was also shown
that the Vandermonde determinant at the critical point $\mu$ is given by
$$\D (X^{(\mu )}_r)=\sum _{w\in W} (-1)^w e^{-2\pi i w(\rho )\cdot \mu}.$$ 
Thus,
using the expression in \gw\ for the modular transformation matrix $S_{\mu
\lambda}$, we have
$$h^o=(-1)^{N(N-1)/2}\big|{W\over (k+N)R}\big|\D ^{-2}$$
where, as usual, $h^o(X^{(\mu )}_r)=1/S^2_{o,\mu}$.  Relation \hw\ is thus the
expected \sunh .  We have thus explicitly
verified relation \th .  The Verlinde dimensions are then expressed as TLG
residues as in \soakres .

\newsec{Simple Cosets}

We can give a LG description of the unitary $N=0$ minimal models
$SU(2)_1\times SU(2)_{m-2}/SU(2)_{m-1}$, constructing the coset along the
lines of \ref\GmodH{G. Moore, N. Seiberg, Phys. Lett. B220 (1989) 422}.  The
LG potential for the coset is 
\eqn\mmLG{W(x,y)={1\over \sqrt 2}(W_{m}(x)-W_{m+1}(y))} 
where $W_{k+2}(x)=W_{k+2}(X_1=x,X_2=1)$ is the LG potential for the $SU(2)_k$
fusion rules, $W_{k+2}(2\cos\theta ) =(2/k+2)\cos (k+2)\theta$.  The primary
fields
are given by
\eqn\mmpfs{\p _{r,s}=P_{r-1}(x)P_{s-1}(y) \quad r+s=\ \hbox{even},}
for $r=1,\dots , m-1$ and $s=1,\dots ,m$, where $P_{n}(x)$=
$\p _{n\Lambda _1}=Y_n(x,X_2=1)$; they are the
Chebyshev polynomials $P_n(2\cos\theta)=\sin (n+1)\theta /\sin \theta$.  The
field identifications from the common ${\bf Z}_2$ center of the coset has
allowed us to write \mmLG\ without a field for the $SU(2)_1$.  This center is
generated by $(x,y)\rightarrow (-x,-y)$ so the fields \mmpfs\ with
$r+s$ even are properly invariant.  The Verlinde dimensions are given by
\eqn\mmver{\cor ^o={1\over 2}\<\p _{i_1}\dots \p _{i_n}(\p _{\hat c})^{1-g}\>
_g}
where the right side is computed using the topological LG residue formula
\res\ with the potential \mmLG\ and the field representations \mmpfs\ with $\p
_{\hat c}=\p _{\hat c}^{-1}=\p _{m-1,m}$.  The factor of 1/2 in
\mmver\ is due to the center mentioned above; the sum over all critical points
of $W$ double counts.  We properly obtain, for example, $\< 1\> ^o_{g=0}$=1
and $\< 1 \> ^o_{g=1}$=$m(m-1)/2$. 

\newsec{Counting Instantons with Fusion Residues}

Consider perturbing our original theory $W\N_{N+M+1}$ in $X_1$, the most
relevant operator.  These perturbations were considered in
\ref\toda{P.Fendley,
W. Lerche, S.D. Mathur and N. Warner, Nucl. Phys. B348 (1991) 66} and were
shown to be
integrable;
they describe affine $SU(N+M)$ Toda at an imaginary coupling.
>From \pert\ the perturbed potential is (with a convenient
sign, corresponding to $X_{N+M}=(-1)^{M+1}\beta$) 
\eqn\mrp{W\N _{N+M+1}(X_i) +(-1)^N\beta X_1}
and the chiral primary fields $\p \N _{\mu}$ are unchanged by this
perturbation.  We will show that the topological LG correlation functions of
the $\p \N _{\mu}$ in the perturbed theory \mrp , computed using \res , are
the twisted Verlinde dimensions for a RCFT.  We then discuss a
conjectured connection between these correlation functions and intersections
of surfaces in the moduli space of Grassmannian instantons.

The ideal $dW$=0 of \mrp\ gives the deformed cohomology relation
$X_NY_M=\beta$.  It is convenient to set $\beta =1$ (the factors of $\beta$
can be put back in
at the end by charge (form) counting).  From \grforms\ it is seen that $\p
_{\mu+e_N} =X_N\p _{\mu}$ when $\mu +e_N$ satisfies \nt.  In terms of Young
tableaux, $X_N$ acts on any cohomology element with less than $M$ columns to
give an element with an extra column of $N$ boxes.  Likewise, using \duality ,
it is easily seen that $Y_M=X_N^{-1}$ acts on any cohomology element with less
than $N$ rows in its tableaux to give an element with an extra row of $M$
boxes.  We thus obtain for the product of the element $X_N\equiv s$ with any
element $\p _{\mu}$
\eqn\sact{X_N\times \p _{\mu}=\p _{s\cdot \mu},}
where, for any $\mu=\sum _{l=1}^Nn_le_l$,
\eqn\sactmu{\eqalign{s\cdot \mu &=\mu +e_N\quad \hbox{if}\quad n(\mu )<M\cr
s\cdot \mu &=\sum _{l=1}^Nn_le_{l-1}\quad \hbox{if}\quad n(\mu )=M,\cr}}
where $e_0=0$ and $n(\mu )$ is the sum \nt .  The element $s=X_N$, under the
ring
multiplication, maps any cohomology element $\p _{\mu}$ to the cohomology
element $\p _{s\cdot \mu}$ defined above.  In this way, the cohomology
elements
can be grouped into orbits under the action of $s$.  From \sactmu\ we have
$s^r=\p_{re_N}$ for $r=1\dots M$, $s^r=\p _{Me_{N+M-r}}$ for $r=M\dots M+N$. 
In particular, $s^{N+M}=1$.  Since $s$ has a ${\bf Z}_{N+M}$=$SU(N+M)_1$
action, the
number of elements in the different $s$ orbits will be divisors of $N$+$M$. 
This $SU(N+M)_1$ symmetry of the fusion rules is connected with the
affine $SU(N+M)$ description of this theory.
 
In each $s$ orbit we can pick a convenient element $\p _{[\mu ]}$ to be the
orbit base point, the other elements in the orbit are of the form $s^r\p
_{[\mu ]}$.  Since the cohomology ring is commutative and associative, ring
products of the elements $\p _{[\mu ]}$, one for each $s$ orbit, suffice to
give the ring multiplication of any two elements:
$$\p _{\mu}\times \p _{\nu}=s^{r+p}\times (\p _{[\mu ]}\times \p _{[\nu ]}).$$
It is convenient to choose our $s$ orbit representatives $[\mu ]$ to have
$n_N=0$ (from \sactmu\ we can clearly make this choice).

If we were to mod out by the $s$ action, we would set $s=X_N=(Y_M)^{-1}=1$. 
Setting $X_N=1$ reduces this theory to the theory discussed in \g\ and in the
last section.  The product of our $n_N=0$ orbit representatives, upon setting
$X_N=1$, thus satisfy the $SU(N)_M$ fusion rules.  The ring structure
without setting $s=1$ is therefore a simple modification of the $SU(N)_M$
fusion
rules.  The product of any two elements, both with $n_N=0$, is given by:
\eqn\mrpfr{\p _{\mu _1}\times \p _{\mu _2}=\sum _{\mu _3}N_{\mu _1 \mu
_2}^{\mu _3}s^{(r(\mu _1)+r(\mu _2)-r(\mu _3))/N}\p _{\mu _3}}
where $N_{\mu _1 \mu _2}^{\mu _3}$ are the fusion rules of $SU(N)_M$, $r(\mu
)$ is the form number of $\p _{\mu}$, $r(\mu )=\sum _{l=1}^{N-1}ln_l$, the
powers of $s$ follow from form counting, and $\p _{Me_1}=s^{-1}$.  In terms of
$SU(N)_M$ Young tableaux multiplication, the idea is to replace each column of
$N$ boxes with a $s$ rather than with a one, where $s=(\p _{Me_1})^{-1}$.  It
is thus seen that the
deformed cohomology ring products of \mrp\ are the fusion rules of a
rational conformal field theory.  The $s$ symmetry of the fusion rules is as
discussed in \k .  The topological LG correlation functions \res\ will give
the twisted Verlinde dimensions for this RCFT.

For the $CP ^1$ case ($N$=$M$=1), this ring deformation
is that shown in \w\ to arise in topological sigma models with target space
$CP^1$ where $\beta =e^{i \Theta}$ with $\Theta$ the coupling in the action to
instanton number.  It is conjectured \ref\msri{C. Vafa, MSRI Conf. on Mirror
Symmetry (1991) proceedings} that the deformation \mrp\ in the Kahler form
generally gives the ``quantum deformation'' of the Grassmannian cohomology
ring that arises in topological sigma models with target space the
corresponding Grassmannian manifold and with a coupling as described above in 
the action to instanton number.  The conjecture, in other words, is that
correlation functions \res\ in the perturbed topological LG theory \mrp\ give
information about the
intersections of surfaces in the moduli space of Grassmannian instantons
(some properties of this moduli space were discussed in \ref\sc{S.
Stromme,
``On Parametrized Rational Curves in Grassmann Varieties''
in {\it Space Curves}, F. Ghione et. al. (Eds.), Springer Verlag Lecture Notes
in Mathematics (1987)}).  

It is easy to see from charge counting that the correlation functions of the
chiral ring elements \grforms\ in the topological LG theory with potential
\mrp\ are given by 
\eqn\grcorr{\< \p _{\mu _1} \dots \p _{\mu _n}\> _g= F(\mu _1 \dots \mu
_n;g,k)\beta ^k \quad
\hbox{if}\quad r_T=(N+M)k+(1-g)NM,}
and are zero if the condition on $r_T$ is not satisfied, where $r_T=\sum
_{i=1}^nr(\mu _i)$ is the total form number.
The conjecture is that the coefficients
$F(\mu _1\dots \mu _n;g,k)$ in the above correlation function give the number
of intersections of
$n$ surfaces in the moduli space of degree
$k$ instanton maps $\Phi$ from genus $g$ Riemann surfaces into the
Grassmannian space
$U(N+M)/U(N)U(M)$.  The $i$-th surface is that obtained by requiring the
instanton map $\Phi$ to satisfy $\Phi (z_i)\in \tilde \sigma _{\mu _i}$, where
$z_i$ is the $i$-th arbitrary point on the Riemann surface and where $\tilde
\sigma _{\mu _i}$ is a fixed cycle in the Grassmannian which is Poincare dual
to $\p _{\mu _i}$ (e.g. if $\p _{\mu}$ is the volume form, then $\tilde \sigma
_{\mu}$ is a point in the Grassmannian).  The codimension of the $i$-th
surface is thus $r(\mu _i)$.  The number of
intersections of these surfaces counts the number of instanton maps of a given
degree which satisfy the above conditions; these are the observables in the
topological sigma model.

A check on the
conjecture is that the relation for $r_T$, the sum of the codimensions of the
intersected surfaces, as given by simple charge counting in \grcorr\ in terms
of $k$ and $g$, is the correct moduli space dimension
\nref\sa{Scott Axelrod, communication}\refs{\sc ,\sa}.  It is also encouraging
that, since
the $F(g,k)$ are the twisted Verlinde dimensions for the RCFT discussed above,
the $F(g,k)$ are nonnegative integers.

The correlation functions \grcorr\ are calculated using \res .  It is
sometimes convenient, as discussed in sect. 2, to calculate in terms of the
$q_i$ variables.  In these variables, the critical points are simply given by 
 $q_i=\alpha
\omega
^{n_i}$ where $\omega ^{N+M}=1$ and $\alpha ^{N+M}=(-1)^{N-1}\beta$ with
$n_i=1,\dots N+M$ and $n_i<n_{i+1}$,
$i=1,\dots N$.  The ${\bf Z}_{N+M}$ structure
obtained above is obvious in these variables.  The operator $h$ of \res\ is 
$h=(-1)^{N(N-1)/2}\Delta
^{-2}(N+M)^{N}(\prod
_iq_i)^{N+M-1}$.  
As an example, the interested reader can easily verify our normalization $\<
\p _{\hat c}\> _{g=0}=1$.

We now consider the N=1, i.e. $CP^M$, case.  The
potential \mrp\ is 
\eqn\cp{W={x^{M+2}\over M+2}-\beta x .}
The ring multiplication, setting $\beta=1$, are the expected ${\bf Z}_{M+1}$
rational torus ($SU(M+1)_1$) fusion rules for $s=x$.  The operator $h$ in
\res\ is given by $h=(M+1)x^M=h^0x^M$ and the correlation functions
are the twisted $SU(M+1)_1$ Verlinde dimensions:
\eqn\cpr{\<x^{r_1}\dots x^{r_n}\> _g=(M+1)^g\beta ^k \quad \hbox{if}\quad \sum
_{i=1}^n r_i=(M+1)k+(1-g)M}
and zero otherwise.  We, again, interpret these numbers as representing the
number of intersections of $n$ surfaces in the moduli space of degree $k$
instanton maps from the genus $g$ Riemann surface to $CP^M$ where the $i$-th
surface is obtained by requiring the map to take the $i$-th (arbitrary) point
on the Riemann surface to any point which lies in some fixed cycle Poincare
dual
to $x^{r_i}$.  The above
result agrees with the analysis in \w\ of $CP^1$ instantons and with Axelrod's
analysis of $CP^N$ instantons \sa. 

As another example,
consider $U(5)/U(3)U(2)$, for which
$W=X_1^6/6-X_1^4X_2+3/2X_1^2X_2^2-1/3X_2^3+\beta X_1$.
The $\beta=1$ RCFT ring relations are given by $\p _{e_1}\times \p
_{e_1}=s+s^3\p _{e_1}$ with $s^5=1$ ($\p _{e_1}=X_1$ satisfies the $SU(2)_3$
fusion rules modified as discussed above and $s=X_2$).  The critical points of
$W$ are
thus given by $s=\omega ^p$ and $s^2\p _{e_1}=(1\pm \sqrt{5})/2$ with
$p=0,\dots ,4$ and $\omega ^5=1$.  The operator $h$ in \res\ is as given in
\th\ with $\p _{\hat c}=s^3$ and $h^o=10+5\beta s^2\p _{e_1}$.
Using \res\ we easily
calculate, for example, $\< X_1^r \> _g=F(g,k)\beta
^k$ if $r=5k+6(1-g)$ and zero otherwise with
$$F(g,k)=5(5\sqrt{5})^{g-1}\Big(\big({\sqrt{5}+1\over
2}\big)^{5(k-g+1)}+(-1)^k\big({\sqrt{5}-1\over
2}\big)^{5(k-g+1)}\Big)$$
which is a nonnegative integer 
(note that
$F(g,k+2)=11F(g,k+1)+F(g,k)$ and $F(g+2,k+2)=125F(g,k)$).  It would be
interesting to check the conjecture by directly calculating the intersections
in the moduli space of instantons for this space.  It would also be
interesting to connect this
with the interpretation in terms of the twisted Verlinde dimensions.
\newsec{Conclusions}

We have seen that RCFT fusion rules which can be described as the chiral ring
structure constants of some LG superpotential must have an extra symmetry
(corresponding to spectral flow) and that the correlation functions in the
corresponding topological LG theory give a twisted version of the RCFT's
Verlinde dimensions.  The symmetry \sfbs\ of the RCFT fusion rules also
implies
a symmetry of the deformed superpotential.  In both of the perturbations
considered here, this symmetry appears to be connected with the fact that the
perturbed theories are integrable.  It would
be interesting to understand if there is a general connection between the
integrability of a perturbed superpotential and having its ring give the
fusion rules of a rational conformal field theory.
\vskip .75in

I would like to thank C. Vafa for many valuable suggestions and discussions.
This work was also supported by NSF grant PHY-87-14654.

\listrefs
\bye